# An empirical investigation of the *g*-index for 26 physicists in comparison with the *h*-index, the *A*-index, and the *R*-index

Michael Schreiber

Institut für Physik, Technische Universität Chemnitz, 09107 Chemnitz, Germany

Tel.: (+49) 371 531 21910

Fax: (+49) 371 531 21919

e-mail: schreiber@physik.tu-chemnitz.de

**Abstract**

Hirsch has introduced the *h*-index to quantify an individual's scientific research output by the largest number *h* of a scientist's papers that received at least *h* citations. In order to take into account the highly skewed frequency distribution of citations, Egghe proposed the *g*-index as an improvement of the *h*-index. I have worked out 26 practical cases of physicists from the Institute of Physics at Chemnitz University of Technology and compare the *h* and *g* values. It is demonstrated that the *g*-index discriminates better between different citation patterns. This can also be achieved by evaluating Jin's *A*-index which reflects the average number of citations in the *h*-core and interpreting it in conjunction with the *h*-index. *h* and *A* can be combined into the *R*-index to measure the *h*-core's citation intensity. I have also determined the *A* and *R* values for the 26 data sets. For a better comparison, I utilize interpolated indices. The correlations between the various indices as well as with the total number of papers and the highest citation counts are discussed. The largest Pearson correlation coefficient is found between *g* and *R*. Although the correlation between *g* and *h* is relatively strong, the arrangement of the data set is significantly different, depending on whether they are put into order according to the values of either *h* or *g*.

# 1. Introduction

The *h*-index was introduced by Hirsch (2005) as an easily determinable estimate of the impact of a scientist's cumulative research contribution. It is defined as the highest number of publications of a scientist that received *h* or more citations each, while the other papers have not more than *h* citations each. It has immediately been recognized (Ball, 2005) as an easily computable indicator for a scientist's achievement, because it incorporates both publication quantity and citation quality. Ever since its introduction, the *h*-index has received a lot of attention, as can be seen by the 72 citations that the original paper (Hirsch, 2005) has already accumulated within the first two years which is more than sufficient to enhance Hirsch's Hirsch index. Bornmann and Daniel (2007b) reviewed the research literature on the *h*-index after one year. They have recently compared nine different variants of the *h*-index (Bornmann, Mutz, and Daniel 2008).

One advantage of the *h*-index is its insensitivity to the number of uncited or lowly cited papers. Therefore it discourages the publication of unimportant work, the partitioning into insignificant pieces (compartmentalization of results into the LPUs - least publishable units), or the repeated publication of similar results, as these strategies would enhance the number of papers of an author, but would also be likely to distribute possible citations over more targets, thus reducing the likelihood that a particular paper reaches the *h*-core.

On the other hand, it is a disadvantage that the *h*-index is also insensitive to one or several outstandingly highly cited papers. This is due to the fact that once a paper has entered the *h*-core, i.e., the *h*-defining set, it is no more relevant whether or not it is further cited. As the frequency distribution of citations is usually highly skewed, this insensitivity is considered as a major drawback of the *h*-index (Egghe, 2006c). In order to overcome this problem, Egghe (2006a, 2006b) proposed the *g*-index, defined as the highest number of papers of a scientist that received *g* or more citations on average. This is the highest number of articles that together received $g^2$ or more citations. In this way not only the high number of citations in the top range is taken into account, but also the evolution of the citation counts of highly cited articles is incorporated, which means that an increase of the citation count of the publications in the *g*-core, i.e. the *g*-defining set, can and will eventually lead to an increase of the *g*-index. Another way to measure the citation frequency of the highly cited papers is the calculation of the average number of citations per "meaningful paper" (Podlubny & Kassayova, 2006). Of course, the respective threshold is ambiguous. An obvious choice would be to use the value *h*; this was suggested by Jin (2006) and the result has been labelled *A*-index, because it is an average. Obviously, *A* cannot stand alone, it requires first the determination of *h* and should only be discussed in conjunction with *h*. This conjunction is formalized in the definition of the *R*-index, which owes its name to its definition as the root

$$R = \sqrt{hA} \tag{1}$$

Jin et al. (2007) proposed to measure the citation intensity in the *h*-core by means of the *R*-index, but also cautioned not to use it as a stand-alone index either, but suggested that it should rather be used in conjunction with *h*, too.

Rousseau (2007) has investigated the relation between *h* and *g* for some simple models. Jin et al. (2007) discussed *h*, *A*, *R*, and *g* and the relations between these indices in the power-law model. They also calculated the indices for 14 Price medallists, extending an earlier evaluation of Egghe (2006c), where the author already suggested that it would be interesting to work out more practical cases in other fields. It is purpose of the present paper to present 26 such practical cases in physics. This is an extension of my previous investigation (Schreiber, 2008a) of nine selected data sets. For the present analysis, the data sets were chosen from my own institution without any bias in the selection, so that the results should be representative for an average institute. Moreover, the unbiased choice also allows a meaningful determination of correlation coefficients between the various indices which shall be compared with the values derived by Jin et al. (2007).

**2. Data base**

The data for the subsequent analysis have been compiled in January and February 2007 from the Science Citations Index provided by Thomson Scientific in the ISI in the Web of Science (WoS). A comprehensive analysis of these data with respect to the $h$-index (Schreiber, 2007a) has shown the difficulty to collect the necessary data. In particular, great care has to be taken to exclude homographs, i.e. to identify publications by different authors with the same name and initials. In six out of the 26 cases considered the $h$-index would be wrongly enhanced by 50% or more, if those homographs where overlooked, in one case the discrepancy leads to a factor of 2.73. On the other hand the reverse effect occurred for one data set, because one scientist published under three different names before and after her marriage. This so-called precision problem (Jin et al., 2007) is not so severe for the more often analysed data sets of the most prominent scientists, because usually there are only rather few homographs which reach their outstandingly high values of the $h$-index. For example, for 8 famous physicists investigated by myself (Schreiber, 2008b) the largest enhancement of the $h$-index due to homographs is 10%.

For my previous investigation of the $g$-index (Schreiber, 2008a) I had selected 9 specific data sets for which I expected to find particularly interesting results. To avoid such a prepossessed choice in the present analysis I have included all 26 data sets, which where analysed with respect to the $h$-index (Schreiber, 2007a). As detailed in that publication, this set includes all scientists that have been working as assistants or senior assistants in my group during their research for their habilitation degree or afterwards. Also included are all full and associate professors from the Institute of Physics at my university as well as recently retired colleagues. I label the data sets with superscripts A, B, C…, Z (which restricted the number of considered retired colleagues to four). The size $n$ of the data sets, i.e., the total number of papers $n$ is given in table 1. There I have also listed the number $n_1$ of papers with at least one citation, because usually in the general framework of information production processes only sources (= articles) with at least one item (= citation) are taken into account. In the here considered 26 cases, between 4.1% and 37.3%, on average 20.9% of the papers received no citation at all. Nevertheless, there is a strong correlation between $n_1$ and $n$, with an observed Pearson correlation coefficient of $\rho(n, n_1) = 0.996$, and a Spearman rank-order correlation coefficient of $\rho_s(n, n_1) = 0.948$.

In contrast, but not surprisingly the highest citation count for a paper in every data set, $c(1)$ is much less correlated with the total number $n$ of papers ($\rho_s = 0.253$) and the number $n_1$ of cited papers ($\rho_s = 0.316$). Particularly striking are the high values of $c^E(1) = 279$ in combination with the low number $n^E = 63$ and $c^X(1) = 204$ while $n^X = 35$. In contrast, unusually low values were found, especially $c^D(1) = 73$ for $n^D = 322$ and $c^Q(1) = 24$ for $n^Q = 86$.

As all scientists whose citation records are investigated here are present or former members of the Institute of Physics at my university, one might wonder how strong the dependencies between the 26 data sets are. However, the number of joint publications is comparatively small. For example, in my own case there are only two colleagues with whom I have co-authored seven and three publications, respectively, out of which one contributes to my $h$-index and three to my $g$-index. Of course, I have more joint publications with my present and former assistants and senior assistants, but in most cases the number of joint publications amounts to between 10 and 20% of their entire number of publications and the joint publications contribute at most 11%, on average less than 3% to my $h$-core. In conclusion, I believe that the dependences between the different data sets are not significant.

## 3. Calculation of the indices

The WoS allows an automatic arrangement of the publication list in decreasing order according to the number of citations $c(r)$, where $r$ is the rank attributed to the paper. The $h$-index is readily read off this list as

$$h \leq c(h) \quad \text{while} \quad c(h+1) < h+1, \tag{2}$$

which corresponds to Hirsch's original definition.

Calculating the sum $s(r)$ of the number of citations up to rank $r$,

$$s(r) = \sum_{r'=1}^{r} c(r'). \tag{3}$$

allows us to determine the $g$-index analogously from

$$g^2 \leq s(g) \quad \text{while} \quad s(g+1) < (g+1)^2 \tag{4}$$

which can be rewritten as

$$g \leq \bar{c}(g) \quad \text{while} \quad \bar{c}(g+1) < g+1, \tag{5}$$

utilizing the average number $\bar{c}(r)$ of citations up to rank $r$, i.e.,

$$\bar{c}(r) = \frac{s(r)}{r}. \tag{6}$$

The relation (4) formally expresses the above specification that $g$ is the highest number of articles that together received $g^2$ or more citations. The relation (5) reflects the other verbal definition above, namely that the $g$-index is given by the highest number of papers which received on average $g$ or more citations.

The $A$-index is given by the average number of citations of the publications in the $h$-core, i.e.,

$$A = \bar{c}(h) = s(h)/h. \tag{7}$$

The similarity of (5) and (7) is obvious, both involve the average number of citations, but for different ranks. Consequently, one might expect a strong correlation between $g$ and $A$.

However, at least for the rank $r = h$ this procedure (7) of measuring the citation intensity in the *h*-core is not imperative, one might likewise utilize directly the sum $s(r)$ and determine

$$R^2 = s(h), \qquad (8)$$

which is equivalent to the definition (1) of the *R*-index as can be seen by evaluating (8) with the help of (6) and (7). Now the similarity between (4) and (8) is evident and leads to the expectation that $g$ and $R$ are strongly correlated.

In the special case where $c(r) = h$ for all $r \leq h$, all indices are the same,

$$h = g = A = R. \qquad (9)$$

Jin et al. (2007) have noted that $R = h$ in this case, but stated that this "nice result" does not hold for the sum itself. But if the sum is divided by $h$, i.e., if the average is considered then one obtains $A = s(h)/h = h$, which is as "nice" as $R = h$.

## 4. Results

The citation counts are visualized for the 26 data sets in figure 1, where the summed number $s(r)$ of citations is displayed versus the rank *r*. Thus a constant number of citations would yield a straight line. For some of the data sets such a behaviour is indeed observed for larger values of *r* over a relatively large range of the rank; if the slope of the straight lines is not too small, these indicate the so-called "enduring performers" (Meho, 2007). On the other hand, the "one-hit wonder" is characterized by an unusually high first citation count $c(1)$, most notably for data set X with $c^X(1) = 13.6\ c^X(2)$, in agreement with the above observation at the end of the second section. The colleagues J,V,W also belong into this category, they have $c(1) \geq 2\ c(2)$. Most of the other data sets with a relatively high value of $s(1) = c(1)$ are characterized by rather large values of $c(2)$ and often $c(3)$ as well, leading to a steep initial increase of $s(r)$ in figure 1; these are the colleagues E,I,M and B,P,R, respectively, but also Z although in the last case the actual numbers are quite small. All these cases show a comparatively weak increase of $s(r)$ for large $r$ in figure 1.

As the parabola $r^2$ is also shown in figure 1, one can directly determine the $g$ values which are given by the rank for which the parabola is just below or at the plotted symbol. The derived values of $g$ are compiled in table 1 together with the previously published values of $h$ (Schreiber, 2007a). On average, I obtain $\langle g \rangle = 24.0 \pm 11.8$ and $\langle h \rangle = 14.9 \pm 6.8$. (The uncertainty in these values is the standard deviation of the mean.) As expected (Egghe, 2006a) the $g$-index yields a larger variance in comparison with the Hirsch index $h$ and it therefore enables a better discrimination between the data sets, although in the present study there are eight pairs of data sets with coinciding $g$-values. But for the $h$-index besides two such pairs there are two h-values that occur three times and one even four times. This makes the order of the $h$-list rather ambiguous. But the clustering of the $h$-values around the median $h_{med} = 14$ and the tails extending to $h_{min} = 5$ and $h_{max} = 39$ agree with expectations for a random sample. Likewise, the $g$-values are clustered around the median $g_{med} = 22$, with longer tails extending from $g_{min} = 10$ to $g_{max} = 67$. A significant rearrangement occurs, when the data sets are sorted according to these $g$-values. Whether this rearrangement is adequate, is of course a matter of interpretation. But in my opinion it is reasonable, as will be discussed below.

Comparing the derived values of the indices with the status of the scientists as given in table 1, it is not surprising, that full professors take the first 3 positions in the $h$-sorted list and the first 4 positions in the $g$-sorted list. But the fourth and fifth position, respectively, is occupied by an associate professor (data set D). Relatively high in the $h$-sorted list are the assistants I and J, and the scientist I even advances to the 6th position in the $g$-sorted list. On the other hand three full professors end up among the last 5 positions in the $h$-sorted list and the last 8 positions in the $g$-sorted list. This shows that even with a comparatively small number of publications and a relatively weak impact as measured by the citation count, it is possible to obtain a full professorship. In order to avoid misinterpretations, I want to point out that one of these scientists obtained the full professorship not at Chemnitz University of Technology.

In table 1 the citation count $c(g+1)$ of the first paper that does not belong to the $g$-core is also included. This is interesting in comparison with $2g+1$ which is the number of additional citations which are needed to increase the index from $g$ to $g+1$. Evidently, the consideration of the citations of one more paper is by far not enough to increase the index, in addition the overall citation counts of the other papers in the $g$-core have to increase substantially. An extreme case in this respect is data set P with $c^P(g+1) = 3$ for $g = 24$. The values of $c^P(g+1)$ for data sets X and Z are as small or even smaller, but the respective $g$-values are much smaller, so that these cases are not so extreme.

The values of the $A$-index which are also shown in table 1 are significantly larger than the $h$-index, because the average number of citations of the papers in the $h$-core is always much larger than the value of $h$. Also the variance is quite large.

The determined values of the $R$-index, see also table 1, are very similar to the $g$-values. This means that the $h$-core is strongly dominating the rank-frequency distribution, and the citation counts of the further publications in the $g$-core have only a minor influence, because they are already relatively small and do not contribute much. (One would obtain $g = R$ in the extreme situation that $c(r) = 0$ for all $r > h$.) Sorting the list of the data sets according to the $R$-index yields only very small rearrangements in comparison with the order that was determined from the $g$-values.

## 5. Interpolation of the indices

Obviously, the $R$-values are real numbers, and the values of $A$ are rational numbers, while by definition $h$ and $g$ take integer values only. As some of the data sets yield rather small values of the $h$-index, 14 of the 26 data sets cannot be unambiguously put into order according to the

*h*-values, because their *h*-indices are not unique. Unexpectedly, the respective unambiguity is even larger for the *g*-index, for which 16 values in table 1 are not unique. This is surprising, because one would expect a better distinction in view of the larger values of the *g*-index and their larger variance. This difficulty can be remedied by a simple generalization of the above definitions, which also solves another minor problem, namely that the definitions of *h* and *g* are unfavourable, when $c(h) > h$ or $s(g) > g^2$, respectively. Although it will make only a small difference in the individual *h*- and *g*-values, it appears adequate to discriminate somewhat more, which is easily possible by a piecewise linear interpolation of the rank-frequency function

$$\tilde{c}(x) = c(r) + (x - r)(c(r+1) - c(r)) \tag{10}$$

between *r* and *r*+1, and then calculating an interpolated Hirsch index $\tilde{h}$ as

$$\tilde{c}(\tilde{h}) = \tilde{h}. \tag{11}$$

This generalization has already been suggested by Rousseau (2007). It makes a difference only when $c(h) > h$, i.e., when the equality in relation (2) does not hold. This was indeed found in 11 of the 26 cases in the present investigation. For the respective generalization of the *g*-index, one can utilize the piecewise linear interpolation of the sum (3)

$$\tilde{s}(x) = s(r) + (x - r)(s(r+1) - s(r)) = s(r) + (x - r)c(r+1) \tag{12}$$

between *r* and *r*+1. Then the interpolated *g*-index $\tilde{g}$ is obtained by setting

$$\tilde{s}(\tilde{g}) = \tilde{g}^2. \tag{13}$$

$\tilde{g}$ differs from *g* when $s(g) > g^2$, i.e., when the equality in relation (4) is not fulfilled. Due to the large numbers involved, usually $s(g) \neq g^2$; in the present study this occurred in 25 of the 26 cases.

The generalization increases the *h* and *g* values by less than one, and the usual integer results can be obtained by truncating the interpolated values, i.e., by the floor function of $\tilde{h}$ and $\tilde{g}$, respectively:

$$h = \lfloor \tilde{h} \rfloor, \tag{14}$$

$$g = \lfloor \tilde{g} \rfloor. \tag{15}$$

For the sake of consistency, the *A*-index and the *R*-index should be evaluated for the interpolated index $\tilde{h}$, i.e.,

$$\tilde{A} = \tilde{c}(\tilde{h}) = \frac{\tilde{s}(\tilde{h})}{\tilde{h}}, \tag{16}$$

$$\tilde{R} = \sqrt{\tilde{s}(\tilde{h})} = \sqrt{\tilde{h}\tilde{A}}. \tag{17}$$

These generalized indices are also bounded by the original indices, namely

$$\tilde{A} \leq A, \tag{18}$$

$$\tilde{R} \geq R. \tag{19}$$

This can be easily proven:

Obviously $\tilde{A} = A$ if $\tilde{h} = h$; and $\tilde{A} < A$ if $\tilde{h} > h$, because $\tilde{c}(x)$ is a decreasing function and it is strictly decreasing between *h* and *h*+1 if $\tilde{h} > h$. Likewise, $\tilde{R} = R$ if $\tilde{h} = h$; and usually $\tilde{R} > R$ if $\tilde{h} > h$, because $\tilde{s}(x)$ is an increasing function and it is strictly increasing between *h* and *h*+1 except in the extremely unlikely case that *c*(*h*+1) = 0 (but even then we still have $\tilde{R} = R$).

In the special case where *c(r)* = *h* for all $r \leq h$, all indices are the same, interpolated or not,

$$\tilde{h} = \tilde{g} = \tilde{A} = \tilde{R} = h = g = A = R. \tag{20}$$

## 6. Results for the interpolated indices and their ratios

Table 2 contains the derived values of $\tilde{h}$ which after truncation correspond with the *h*-values in table 1. The data sets have been put into order according to these values of $\tilde{h}$. Only for two pairs (M,N and P,Q) the $\tilde{h}$-values coincide, and in these cases I have decided the order by the values of the *g*-index.

The $\tilde{g}$-values can be read off figure 1, when the intersections of the parabola with the linear interpolation between the symbols are determined. The effect of the generalization of *g* to $\tilde{g}$ can be exemplified comparing the cases Q and T for which the discrete evaluation of *g* yields the same value $g^Q = g^T = 15$ while the interpolation results in $\tilde{g}^Q = 15.9$ in contrast to $\tilde{g}^T = 15.1$, which reflects the already significant separation of the data in the figure. A similar observation can be made for data sets O and R, for L and N, as well as for M and P. On the other hand the critical value $g = 22$ separates the symbols for data sets K and N which are quite close so that the interpolated values $\tilde{g}$ are quite close, $\tilde{g}^N = \tilde{g}^K + 0.09$, although the integer values differ, $g^N = g^K + 1$. An analogous behaviour is found for data sets G and M, for Y and Z, as well as for C and E. The derived values of $\tilde{g}$ are listed in table 2.

Egghe (2006c) noted that the ratio *g/h* could be an interesting measure, because this relative increase indicates how strongly skewed the frequency distribution of citations is. In the present investigation this can be clearly observed, see table 2, with a high value of $\tilde{g}^X / \tilde{h}^X = 2.27$ in striking contrast to $\tilde{g}^Q / \tilde{h}^Q = 1.22$. The ratio is also quite low for data sets Y, G, T, and O, while it is well above the average of $\langle \tilde{g} / \tilde{h} \rangle = 1.62$ for the data sets E, P, I, and Z, too. If the most-cited paper is excluded from data set X (the one-hit wonder), one still obtains $\tilde{h}^{\overline{X}} = 8.0$, but $\tilde{g}^{\overline{X}} = 9.9$, and thus $\tilde{g}^{\overline{X}} / \tilde{h}^{\overline{X}} = 1.23$. This demonstrates the extreme influence of the exceptionally high value of $c^X(1)$ on the indices.

Jin et al. (2007) suggested, that the ratio *R/h* might be an interesting indicator in its own right, solving the problem that as a stand-alone index $R$ may be overly sensitive to one article receiving an extremely high number of citations. As discussed above, in the present investigations data sets E and X are outstanding in this respect. It is therefore not surprising, that their $R$- and $\tilde{R}$-values are relatively high. But as table 2 shows, this is true for $\tilde{R}/\tilde{h}$ as well. Looking at the other $\tilde{R}/\tilde{h}$-results, the rather high value for data set P is conspicuous, indicating a strongly skewed citation record which was not so obvious from $c^P(1) = 108$, but could have been expected because $c^P(g+1) = 3$ is extremely small.

In general, the ratio $\tilde{R}/\tilde{h}$ is very similar to the discussed ratio $\tilde{g}/\tilde{h}$, see also table 2. On average, $\langle \tilde{R}/\tilde{h} \rangle = 1.48 \pm 0.20$ and $\langle \tilde{g}/\tilde{h} \rangle = 1.62 \pm 0.21$. A very strong agreement between $g$ and $R$ was already observed by Jin et al. (2007) and can be quantified by the ratio $\tilde{R}/\tilde{g}$, which fluctuates only a little bit around the average value of $\langle \tilde{R}/\tilde{g} \rangle = 0.913 \pm 0.012$. For the ratio $R/g$, i.e., without interpolation I obtained an average value of $\langle R/g \rangle = 0.929 \pm 0.024$. I consider the difference in the standard deviation, which amounts to a factor of two, as an indication for the superiority of the interpolated indices over the original indices.

The above observations are visualized in figure 2, in which the various indices are compared. Here the data sets are arranged in accordance with their $\tilde{g}$-values. Due to the logarithmic scale one can clearly see the similarity between $\tilde{g}$ and $\tilde{R}$. There is not much difference between them, and the difference is nearly the same for all data sets. It is obvious from this figure, that $\tilde{g}$ (or $\tilde{R}$) leads to a significantly different order of the data sets, as can be seen from the non-monotonic behaviour of the heights of the bars of the lowest histogram reflecting the $\tilde{h}$-values. Of course, the same observation can be made with respect to the values of $\tilde{A}$, because on the logarithmic scale the difference between the height of the $\tilde{R}$- and the $\tilde{h}$-bars is the same as the difference between $\tilde{A}$ and $\tilde{R}$, because $\tilde{A}/\tilde{R} = \tilde{R}/\tilde{h}$ follows from (17). Examining the rearrangement of the data sets in figure 2, one can observe that only 5 positions are unchanged comparing the order with respect to $\tilde{g}$ and the order with respect to $\tilde{h}$, see also table 2. Colleague P advances seven positions, X five positions, I and M three positions, while scientist Q drops five positions, G and K four positions, T three positions in the list. These are quite significant changes. In contrast, comparing the list put into order according to $\tilde{R}$ with that arranged by $\tilde{g}$, 19 positions remain unchanged, 6 scientists advance or drop one position, and only M drops 2 positions. These changes are insignificant.

The last two columns of table 2 show the order which the data sets would have if they were sorted by the ratio $\tilde{g}/\tilde{h}$ or $\tilde{R}/\tilde{h}$. The above discussed cases X, E, and P are of course highest in this order. Large changes in comparison with the $\tilde{h}$-sorted list, as well as with the $\tilde{g}$- or $\tilde{R}$-sorted list can be observed. So I agree with Egghe (2006c) that "a possible interesting measure is *g/h*" and with Jin et al. (2007) that "$R/h$ might be an interesting indicator in its own right", but I do not think that they should be utilized for evaluation purposes.

Rather conspicuous are the discrepancies of more than ten positions in the $\tilde{g}/\tilde{h}$-sorted list for the data set pairs C,E, and G,M, and S,X, as well as Y,Z in comparison with the cumulative citation counts in figure 1, where the respective curves of these pairs show an intersection around or slightly above their *g*-values. This corroborates the expectation that the $\tilde{g}/\tilde{h}$-values discriminate between different citation patterns, here clearly between different slopes of the *s(r)*-curves for similar values of the *g*-index. For the data-set pairs D,I and F,H an equivalent observation can be made: the positions in the $\tilde{g}/\tilde{h}$-sorted list differ by 15 and 10 positions, respectively, while the curves intersect between their *h*- and *g*-values with distinctly different slopes. Even when the intersection occurs below the rank *h*, the effect on the $\tilde{g}/\tilde{h}$-order can be as large as 9 positions, as for the pair S,V.

## 7. Correlations between the indices

To visualize a possible correlation between the 4 indices investigated in the present study, figure 3 shows $\tilde{h}$, $\tilde{R}$, and $\tilde{A}$ in dependence on $\tilde{g}$. Of course, it is not surprising that there is a strong correlation. In contrast, when one compares the number of papers, the number of cited papers, or the maximum number of citations with $\tilde{g}$, as displayed in figure 4, the correlation is much less obvious. In general scientists with a high number of papers usually also have a high number of cited papers and more citations than the average, and thus they usually have higher indices. But it is clear from the strongly scattered data points in figure 4 as compared to figure 3 that the correlations are much weaker.

To quantify these observations, Pearson's correlation coefficients $\rho$ are compiled in table 3. As it is not clear whether the values of the indices follow a normal distribution which should be the case if one calculates Pearson's correlation coefficients, I have also computed Spearman's rank-order correlation coefficients $\rho_s$, see table 4. However, as most values in the literature with which I want to compare are given for Pearson's correlation coefficient, in the following I shall use both coefficients $\rho$ and $\rho_s$.

Following the above discussion, it is not unexpected that the correlation between $\tilde{g}$ and $\tilde{R}$ is very high, with $\rho_s(\tilde{g},\tilde{R}) = 0.997$ and $\rho(\tilde{g},\tilde{R}) = 0.99961$ it is even higher than the values $\rho(g,R) = 0.991 - 0.999$ determined by Jin et al. (2007). The correlation $\rho_s(\tilde{g}/\tilde{h},\tilde{R}/\tilde{h}) = 0.992$ and $\rho(\tilde{g}/\tilde{h},\tilde{R}/\tilde{h}) = 0.996$ is slightly smaller, but still extremely strong, and in very good agreement with the values $\rho(g/h,R/h) = 0.959 - 0.998$ derived by Jin et al. (2007). But the correlation coefficients between the other indices are also high, with values between $\rho(\tilde{g},\tilde{A}) = 0.971$ and $\rho(\tilde{g},\tilde{h}) = 0.975$. The only exception is the relatively weak correlation $\rho(\tilde{h},\tilde{A}) = 0.895$ between $\tilde{h}$ and $\tilde{A}$, which can be easily explained, because a high value of $\tilde{h}$ in comparison with $\tilde{R}$ means a low value of $\tilde{A}$ in comparison with $\tilde{R}$ and vice versa, by definition of $\tilde{R}$, see (17). The rank-order correlation coefficients are slightly smaller, between $\rho_s(\tilde{g},\tilde{h}) = 0.931$ and $\rho_s(\tilde{g},\tilde{A}) = 0.938$, again with the exception $\rho_s(\tilde{h},\tilde{A}) = 0.793$. On the basis of a stochastic model, Burrell (2007) has recently suggested a proportionality between $\tilde{h}$ and $\tilde{A}$. The current data show this behaviour, but as the correlation coefficients indicate, the proportionality between $\tilde{h}$ and $\tilde{g}$, as well as between $\tilde{g}$ and $\tilde{A}$ is much more pronounced.

In general, the correlation of the indices with the total number $n$ of papers is significantly lower, see also tables 3 and 4. It is highest for $\tilde{h}$ and lowest for $\tilde{A}$, which is not surprising, because $\tilde{A}$ is most strongly influenced by the number of citations so that this relatively low value means, that the correlation between visibility and publication quantity is not so strong. Correspondingly the correlation of all the indices with the number $n_1$ of cited papers is somewhat larger than that with the total number of papers. Considering the highest citation count $c(1)$, its rank-order correlation with $\tilde{A}$ is of course rather high, namely 0.912. On the other hand, the rank-order correlation between $c(1)$ and $\tilde{h}$ is comparatively weak ($\rho_s = 0.618$) in accordance with the above discussion in which some data sets are characterized as the one-hit or two-hit wonders in contrast to the enduring performers.

One should be aware, however, that even the "comparatively weak" correlations are highly significant. Significant correlations $\rho_s(h,n) \approx 0.84 – 0.94$ between $n$ and $h$ have also been reported recently by Bornmann and Daniel (2007a) in contrast to Vinkler (2007), who obtained $\rho(h,n) = 0.40$ which is not significant. On the other hand, van Raan (2006) derived a moderate correlation $\rho(\log h, \log n) = 0.697$ for the logarithmized values.

## 8. Further discussion and summary

In this investigation I have analysed the citation records of 26 physicists of the Institute of Physics at Chemnitz University of Technology. I assume that the unbiased choice of these scientists has yielded a representative sample for an average institute. As expected, the $g$-index allows for a better discrimination between the data sets and yields some rearrangement of the order. The rearrangements can be traced to different individual citation patterns, in particular distinguishing between one-hit wonders and enduring performers: the one-hit wonders advance in the $g$-sorted list. In my opinion, this makes the $g$-index more suitable than the $h$-index to characterize the overall impact of the publications of a scientist. Especially for not-so-prominent scientists the small values of $h$ do not allow for a reasonable distinction between the data sets. This situation is also improved by utilizing the $g$-index. Nevertheless, one needs to be aware that small differences in the resulting values should not be overinterpreted.

This means that certainly the differences between the original integer values and the interpolated values should not be used to decide, whether one scientist is better than the other. But for the purpose of the present analysis, I found it helpful to smoothen the step functions which arise from the integer definitions of $h$ and $g$. Of course, the effect is small, but it also applies to the derived indices $A$ and $R$, which take rational and real numbers in the original definition anyway. Obviously, the relative influence of the interpolation will be stronger for smaller values of the indices, therefore one should rather utilize the generalized indices when comparing many data sets with very small values of $h$ and $g$, like in the study of Bornmann, Mutz, and Daniel (2008). Likewise the generalization will be much more important when investigating the time dependence of the indices with a limited time window which also leads to small values of the indices. Such an analysis of the current 26 data sets is left for the future work.

Psychologically, the interpolated indices have the advantage, that one does not have to wait so long to see one's index growing. For the $\tilde{h}$-index this is only a minor point, because the growth depends on the number of $c(h)$ and $c(h+1)$ only. In contrast, for the $\tilde{g}$-index every citation of papers in the entire $g$-core counts and leads to an albeit small increase of the $\tilde{g}$-value. Therefore people can "enjoy" an increase of the popularity of their publications much more often.

In this context I reiterate my deliberations (Schreiber 2007a, b, 2008a) that the influence of self-citations can and will in several cases strongly influence the citation records and should therefore be eliminated from the data base, as also demanded by Vinkler (2007), in order to allow a meaningful comparison between the impact of the publications of different scientists. I admit, that this is very difficult and requires a very labour-intensive analysis, but in my opinion this effort is necessary if one wants to avoid the criticism, that the calculation of these indices is only meaningless numerology.

The observation that homographs become more significant when lower citation counts have to be considered also means that the precision problem becomes more serious when the $g$-index is determined instead of the $h$-index, because the $g$-core is usually significantly larger than $h$-core. This problem in addition to the larger computational expense for the larger core has led Jin et al. (2007) to the conclusion that the $R$-index is more advantageous. On the other hand, Jin et al. (2007) have also listed as one disadvantage of the $h$-index, that "it is only useful for comparing the better scientists in a field. It does not discriminate among average scientists." Although in my opinion this is exaggerated, it is certainly true, that the $g$-index allows for a better discrimination among average scientists as demonstrated in the present investigation. The $R$-index does a similarly good job in distinguishing average scientists. In my opinion, the $g$-index is slightly better, because it favours the "enduring performer" with a higher citation count for publications with a rank beyond the value of *h*. However, this may not be worth the additional computational effort and the enlarged precision problem.

But one argument remains in favour of the *g*-index: it is just more elegant, because it is just one number that needs to be determined. Returning to the argumentation in the introduction what would be the best threshold to calculate the average number of citations per "meaningful paper", the definition of the *g*-index removes the ambiguity, because it self-consistently determines the number of meaningful papers as equal to the average number of citations per meaningful paper. Whether it is more or less elegant, however, does not help to answer the fundamental question concerning the usage of *h* or *g*, namely whether it is acceptable "to reduce a lifetime's work to a number" (Kelly & Jennions, 2006). I doubt it, but I leave the further contemplation to the reader.

Table 1. Characteristics of the 26 data sets analysed in the present study. The first column labels the data sets, the next column gives the status of the scientist where 1 indicates an assistant or assistant professor position (C2- or W1-professorship in Germany), 2 an associate professor position (C3 or W2) and 3 a full professorship (C4 or W3). The following columns show the total number $n$ of publications, the number $n_1$ of publications which received at least one citation, the highest citation count $c(1)$, the citation count $c(g+1)$ of the first paper beyond the $g$-core, and the indices $h, g, A, R$.

| data set | status | $n$ | $n_1$ | $c(1)$ | $c(g+1)$ | $h$ | $g$ | $A$ | $R$ |
|---|---|---|---|---|---|---|---|---|---|
| A | 3 | 290 | 250 | 457 | 22 | 39 | 67 | 93.9 | 60.5 |
| B | 3 | 270 | 214 | 182 | 18 | 27 | 45 | 62.6 | 41.1 |
| C | 3 | 126 | 103 | 129 | 15 | 23 | 36 | 47.3 | 33.0 |
| D | 2 | 322 | 259 | 73 | 17 | 20 | 29 | 35.5 | 26.6 |
| E | 3 | 63 | 57 | 279 | 6 | 19 | 37 | 62.4 | 34.4 |
| F | 2 | 131 | 107 | 53 | 14 | 18 | 26 | 32.2 | 24.1 |
| G | 2 | 49 | 47 | 57 | 9 | 17 | 23 | 28.4 | 22.0 |
| H | 3 | 70 | 47 | 70 | 7 | 16 | 26 | 35.9 | 24.0 |
| I | 1 | 65 | 53 | 149 | 6 | 15 | 28 | 46.1 | 26.3 |
| J | 1 | 51 | 32 | 112 | 5 | 15 | 23 | 32.1 | 21.9 |
| K | 2 | 79 | 56 | 55 | 10 | 14 | 21 | 27.7 | 19.7 |
| L | 2 | 88 | 67 | 64 | 8 | 14 | 22 | 30.6 | 20.7 |
| M | 3 | 70 | 60 | 100 | 8 | 14 | 24 | 34.0 | 21.8 |
| N | 2 | 72 | 61 | 55 | 11 | 14 | 22 | 27.7 | 19.7 |
| O | 2 | 77 | 66 | 47 | 9 | 13 | 19 | 22.8 | 17.2 |
| P | 3 | 47 | 37 | 108 | 3 | 13 | 24 | 41.5 | 23.2 |
| Q | 1 | 86 | 59 | 24 | 10 | 13 | 15 | 17.1 | 14.9 |
| R | 1 | 46 | 37 | 53 | 8 | 12 | 19 | 27.0 | 18.0 |
| S | 2 | 61 | 48 | 40 | 7 | 12 | 18 | 22.8 | 16.6 |
| T | 2 | 78 | 56 | 31 | 9 | 10 | 15 | 18.0 | 13.4 |
| U | 2 | 44 | 34 | 41 | 7 | 10 | 17 | 23.7 | 15.4 |
| V | 3 | 60 | 49 | 79 | 6 | 10 | 17 | 24.4 | 15.6 |
| W | 3 | 53 | 37 | 42 | 7 | 9 | 13 | 15.6 | 11.8 |
| X | 3 | 35 | 29 | 204 | 3 | 8 | 18 | 35.1 | 16.8 |
| Y | 2 | 25 | 19 | 19 | 5 | 7 | 9 | 11.0 | 8.8 |
| Z | 2 | 15 | 12 | 25 | 2 | 5 | 10 | 17.0 | 9.2 |

Table 2. The interpolated indices $\tilde{h}$, $\tilde{g}$, $\tilde{A}$, $\tilde{R}$, and the ratios between $\tilde{h}$, $\tilde{g}$, and $\tilde{R}$ for the 26 data sets in table 1, as well as the ranks which the data sets would hold when the list were sorted according to $\tilde{g}$, $\tilde{R}$, $\tilde{g}/\tilde{h}$, or $\tilde{R}/\tilde{h}$, respectively.

| data set | $\tilde{h}$ | $\tilde{g}$ | $\tilde{A}$ | $\tilde{R}$ | $\tilde{g}/\tilde{h}$ | $\tilde{R}/\tilde{h}$ | $\tilde{R}/\tilde{g}$ | $r[\tilde{g}]$ | $r[\tilde{R}]$ | $r[\tilde{g}/\tilde{h}]$ | $r[\tilde{R}/\tilde{h}]$ |
|---|---|---|---|---|---|---|---|---|---|---|---|
| A | 39.0 | 67.1 | 93.9 | 60.5 | 1.72 | 1.55 | 0.90 | 1 | 1 | 7 | 7 |
| B | 27.5 | 45.6 | 62.0 | 41.3 | 1.66 | 1.50 | 0.90 | 2 | 2 | 9 | 9 |
| C | 23.0 | 36.7 | 47.3 | 33.0 | 1.60 | 1.43 | 0.90 | 4 | 4 | 13 | 15 |
| D | 20.0 | 29.8 | 35.5 | 26.6 | 1.49 | 1.33 | 0.90 | 5 | 5 | 19 | 20 |
| E | 19.3 | 37.2 | 61.7 | 34.5 | 1.92 | 1.79 | 0.93 | 3 | 3 | 2 | 3 |
| F | 18.0 | 26.6 | 32.2 | 24.1 | 1.48 | 1.34 | 0.90 | 7 | 7 | 20 | 19 |
| G | 17.0 | 23.9 | 28.4 | 22.0 | 1.40 | 1.29 | 0.92 | 11 | 10 | 24 | 23 |
| H | 16.0 | 26.2 | 35.9 | 24.0 | 1.64 | 1.50 | 0.91 | 8 | 8 | 10 | 10 |
| I | 15.3 | 28.8 | 45.4 | 26.4 | 1.88 | 1.72 | 0.92 | 6 | 6 | 4 | 5 |
| J | 15.0 | 23.6 | 32.1 | 21.9 | 1.57 | 1.46 | 0.93 | 12 | 11 | 16 | 13 |
| K | 14.5 | 22.0 | 27.2 | 19.9 | 1.52 | 1.37 | 0.90 | 15 | 14 | 18 | 18 |
| L | 14.4 | 22.7 | 30.1 | 20.8 | 1.58 | 1.44 | 0.92 | 13 | 13 | 14 | 14 |
| M | 14.0 | 24.1 | 34.0 | 21.8 | 1.72 | 1.56 | 0.90 | 10 | 12 | 6 | 6 |
| N | 14.0 | 22.1 | 27.7 | 19.7 | 1.58 | 1.41 | 0.89 | 14 | 15 | 15 | 16 |
| O | 13.3 | 19.1 | 22.6 | 17.3 | 1.43 | 1.30 | 0.91 | 17 | 17 | 22 | 22 |
| P | 13.0 | 24.7 | 41.5 | 23.2 | 1.90 | 1.79 | 0.94 | 9 | 9 | 3 | 2 |
| Q | 13.0 | 15.9 | 17.1 | 14.9 | 1.22 | 1.15 | 0.94 | 22 | 22 | 26 | 26 |
| R | 12.3 | 19.8 | 26.6 | 18.1 | 1.60 | 1.47 | 0.92 | 16 | 16 | 12 | 12 |
| S | 12.0 | 18.2 | 22.8 | 16.6 | 1.52 | 1.38 | 0.91 | 18 | 19 | 17 | 17 |
| T | 10.7 | 15.1 | 17.5 | 13.7 | 1.42 | 1.28 | 0.91 | 23 | 23 | 23 | 24 |
| U | 10.5 | 17.2 | 23.0 | 15.6 | 1.64 | 1.48 | 0.90 | 21 | 21 | 11 | 11 |
| V | 10.3 | 17.2 | 24.0 | 15.7 | 1.68 | 1.53 | 0.91 | 20 | 20 | 8 | 8 |
| W | 9.0 | 13.2 | 15.6 | 11.8 | 1.46 | 1.31 | 0.90 | 24 | 24 | 21 | 21 |
| X | 8.0 | 18.2 | 35.1 | 16.8 | 2.27 | 2.10 | 0.92 | 19 | 18 | 1 | 1 |
| Y | 7.0 | 9.5 | 11.0 | 8.8 | 1.36 | 1.25 | 0.92 | 26 | 26 | 25 | 25 |
| Z | 5.3 | 10.0 | 16.2 | 9.3 | 1.88 | 1.74 | 0.93 | 25 | 25 | 5 | 4 |

Table 3. Pearson's correlation coefficients $\rho$ for $\tilde{g}$ and $\tilde{h}$ with the other coefficients investigated in this study, as well as with the total number of publications, the number of cited publications, and the highest citation count.

|   | $\tilde{h}$ | $\tilde{g}$ | $\tilde{R}$ | $\tilde{A}$ | $n$ | $n_1$ | $c(1)$ |
|---|---|---|---|---|---|---|---|
| $\tilde{h}$ | 1 | 0.975 | 0.972 | 0.895 | 0.806 | 0.827 | 0.757 |
| $\tilde{g}$ | 0.975 | 1 | 1.000 | 0.971 | 0.744 | 0.772 | 0.863 |

Table 4. Same as table 3, but for Spearman's rank-order correlation coefficient $\rho_s$.

|   | $\tilde{h}$ | $\tilde{g}$ | $\tilde{R}$ | $\tilde{A}$ | $n$ | $n_1$ | $c(1)$ |
|---|---|---|---|---|---|---|---|
| $\tilde{h}$ | 1 | 0.931 | 0.936 | 0.793 | 0.726 | 0.731 | 0.618 |
| $\tilde{g}$ | 0.931 | 1 | 0.997 | 0.938 | 0.580 | 0.636 | 0.790 |

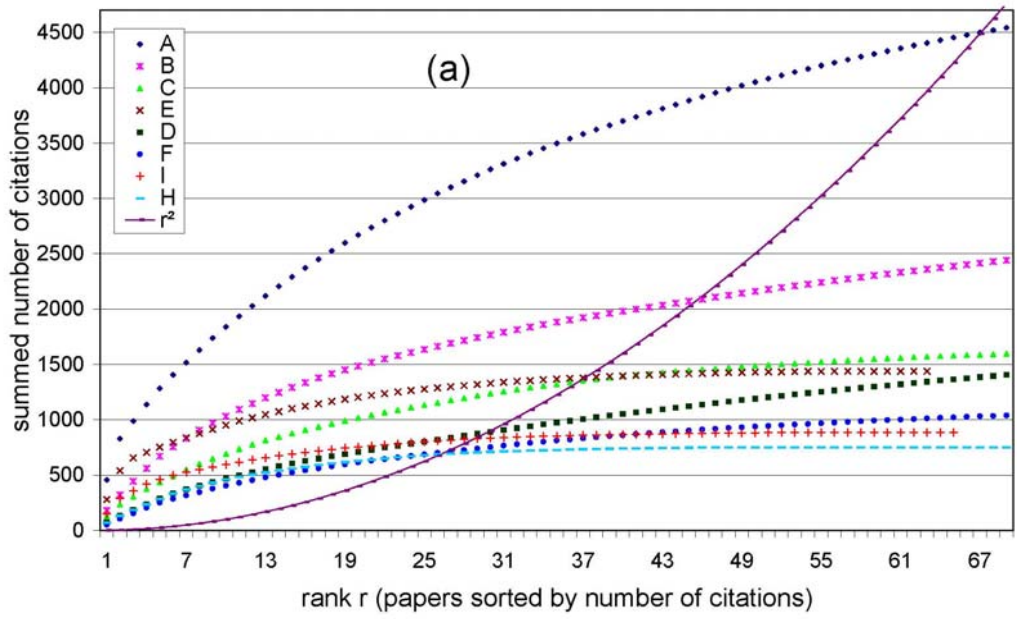

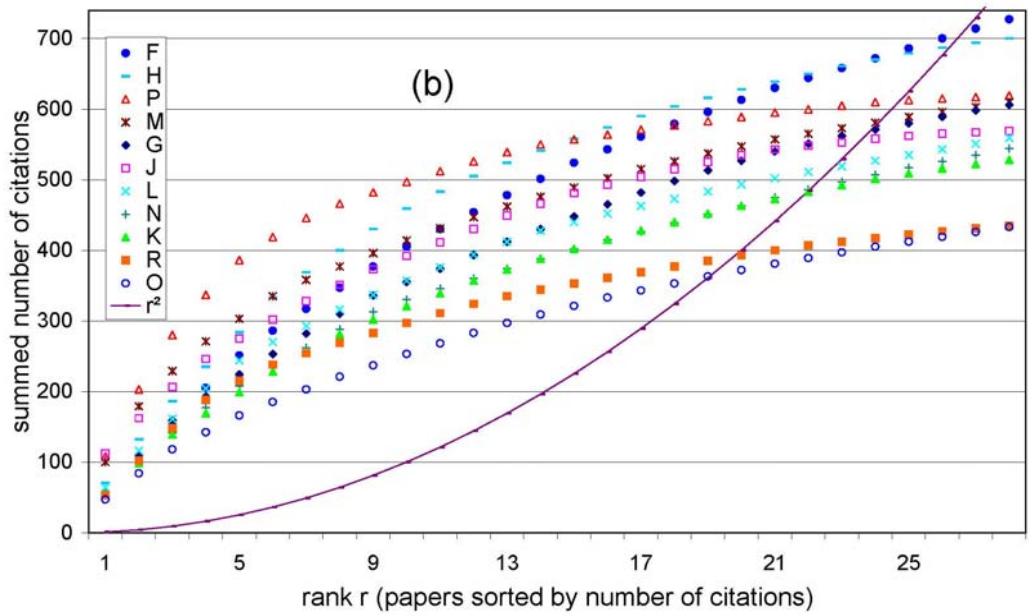

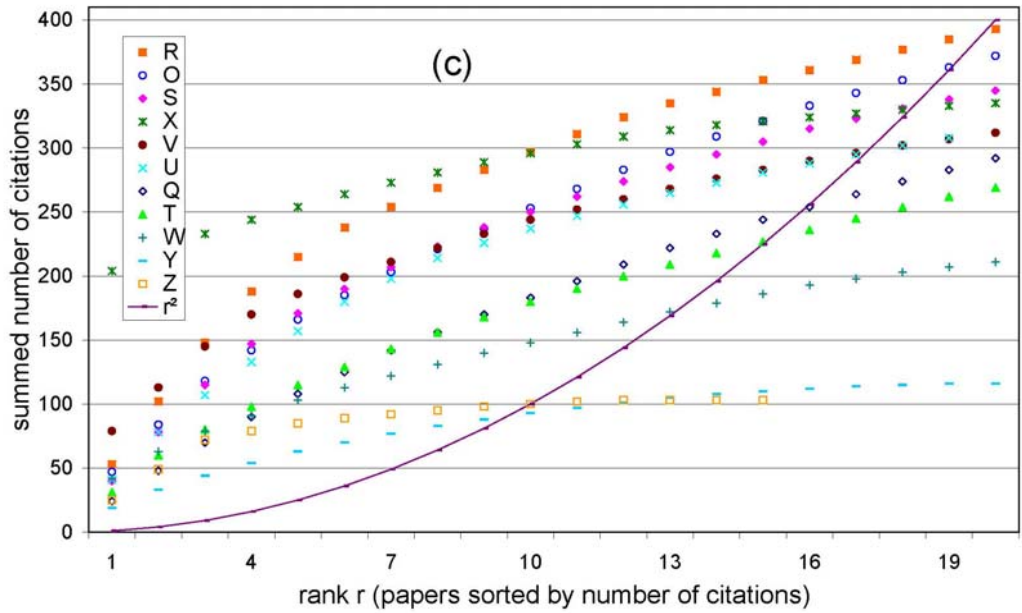

Figure 1. Summed number $s(r)$ of citations of the $r$ most cited papers in each data set. The data set labels A,B,C, …,Z in the legend are arranged according to the citation counts for large $r$, i.e., on the right ends of the plots. Missing symbols for data sets E, I, and Z are due to the fact that these publication lists are exhausted. For better comparison between the three panels, the data sets F and H are included in panels (a) and (b), and the data sets O and T are included in panels (b) and (c). The parabola $r^2$ is also shown.

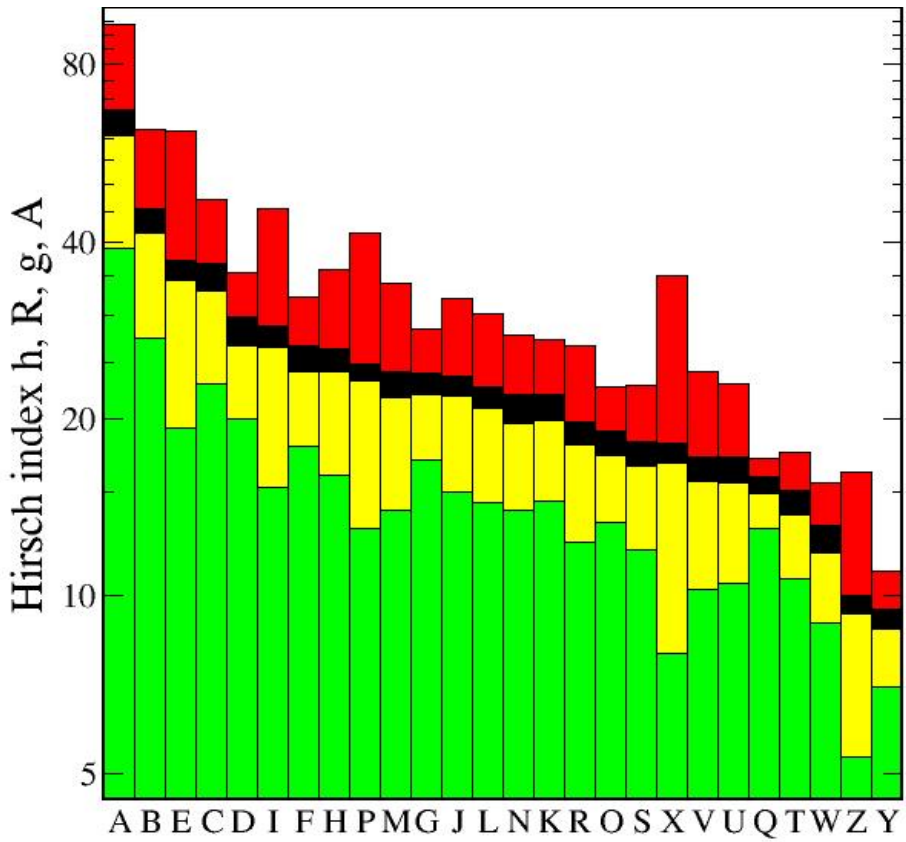

Figure 2. Different indices for the 26 investigated data sets. From top to bottom: $\tilde{A}$ (dark grey/red), $\tilde{g}$ (black), $\tilde{R}$ (light grey/yellow), $\tilde{h}$ (medium grey/green). The data sets are put into order according to the $\tilde{g}$-index, as indicated at the horizontal axes, where the letters are not in alphabetical order in contrast to the sequence in table 1 determined by the original index $\tilde{h}$. Note the logarithmic scale.

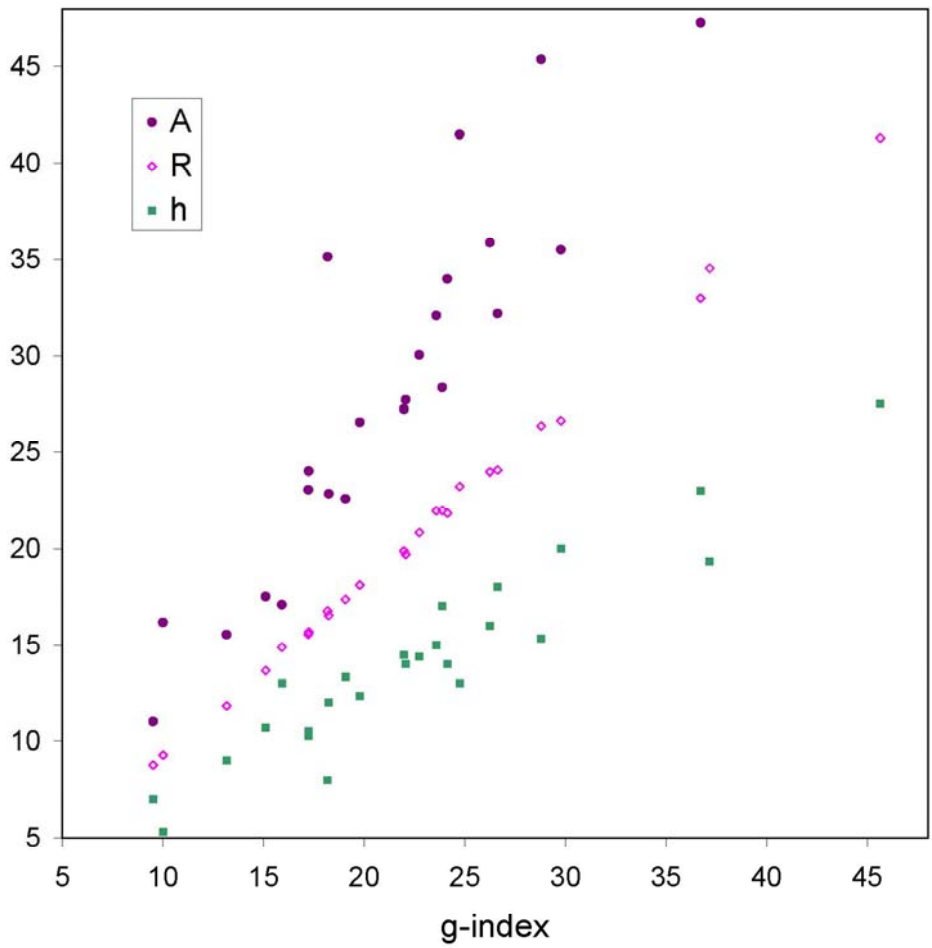

Figure 3. Scatter plot of $\tilde{A}$, $\tilde{R}$ and $\tilde{h}$ (from top to bottom) versus $\tilde{g}$. Note that 5 data points lie outside the displayed ranges.

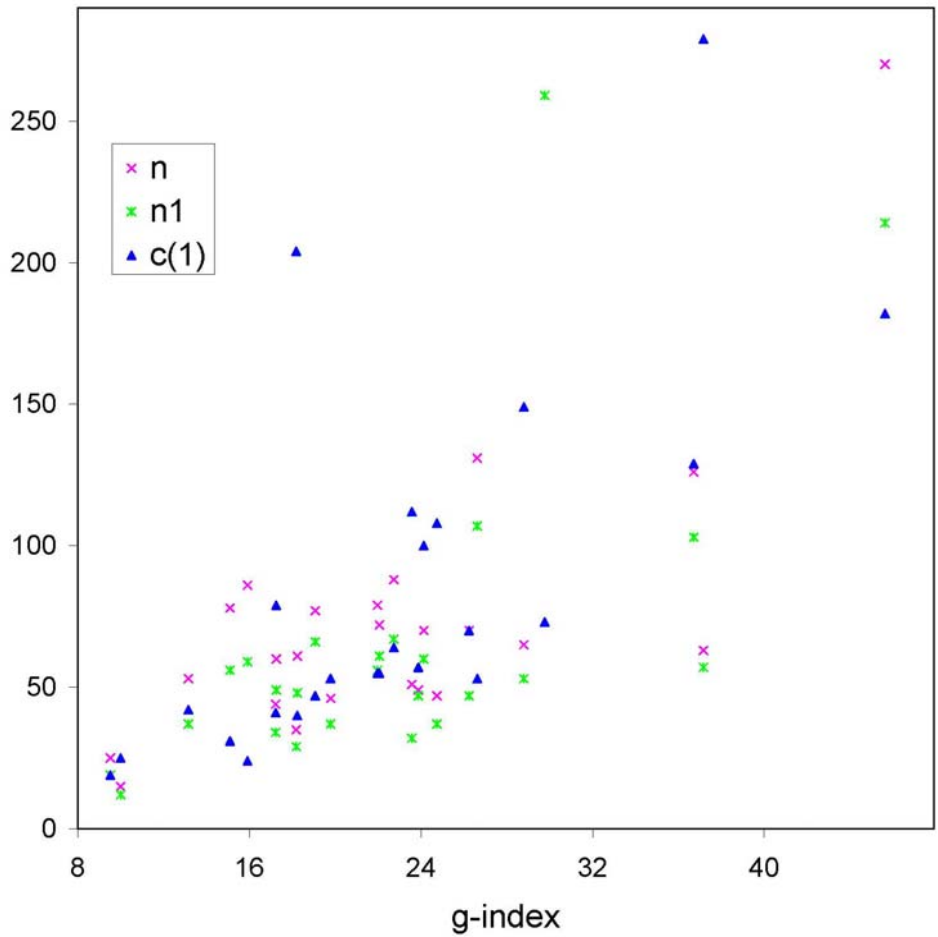

Figure 4. Scatter plot of the total number $n$ of publications, the number $n_1$ of cited publications, and highest citation count $c(1)$ versus $\tilde{g}$. Note that 4 data points lie outside the displayed ranges.